\documentclass[letter,longauth]{aa} 
\usepackage{graphicx}
\usepackage{txfonts}
\usepackage{natbib}
\bibpunct{(}{)}{;}{a}{}{,} 
\usepackage{color}
\newcommand{\um}{$\mu$m}                                 
\newcommand{\lsun}{$L_{\odot}$}                          
\newcommand{\msun}{$M_{\odot}$}
\newcommand{\rsun}{$R_{\odot}$}

\newcommand{\powten}[1]{10$^{#1}$}
\newcommand{\acena}{$\alpha \, {\rm Cen\,A}$}          
\newcommand{\acenb}{$\alpha \, {\rm Cen\,B}$}         
\newcommand{\acen}{$\alpha \, {\rm Cen}$}

\newcommand{\asec}{$^{\prime \prime}$}
\newcommand{\adeg}{$^{\circ}$}

\newcommand{\asecdot}[2]{\mbox{#1$\stackrel {\prime \prime}{_{\bf \cdot}}$#2}}

\begin{document}
   \title{$\alpha$\,Centauri\,A in the far infrared\thanks{Based on observations with {\it Herschel} which is an ESA space observatory with science instruments provided by European-led Principal Investigator consortia and with important participation from NASA. }}

  \subtitle{First measurement of the temperature minimum of a star other than the Sun}

   \author{
   		R. Liseau\inst{1}
	 \and
 		B. Montesinos\inst{2}
	\and
		G. Olofsson\inst{3}
	\and
		G. Bryden\inst{4}
	\and
		J. P. Marshall\inst{5}
	\and
		D. Ardila\inst{6,\,7}
	\and	
		A. Bayo Aran\inst{8,\,9}
	\and
		W. C. Danchi\inst{10}	
	\and	
		C. del Burgo\inst{11}
	\and
		C. Eiroa\inst{5}
	\and
		S. Ertel\inst{12}
	\and
		M. C. W. Fridlund\inst{13}
	\and
		A. V. Krivov\inst{14}
	\and
		G. L. Pilbratt\inst{13}
	\and
		A. Roberge\inst{15}
	\and
		P. Th\'ebault\inst{16}
	\and
		J. Wiegert\inst{1}
	\and
		G. J. White\inst{17,\,18}
	   }
 
   \institute{Department of Earth and Space Sciences, Chalmers University of Technology, Onsala Space Observatory, SE-439 92 Onsala, Sweden,
              \email{rene.liseau@chalmers.se}  
	\and
		Departamento de Astrof\'{\i}sica, Centro de Astrobiolog\'{\i}a (CAB, CSIC-INTA), Apartado 78, 28691 Villanueva de la Ca\~nada, Madrid, Spain
	\and
		Department of Astronomy, Stockholm University, SE-106 91 Stockholm, Sweden
	\and
		Jet Propulsion Laboratory, M/S 169-506, 4800 Oak Grove Drive, Pasadena, CA 91109, USA
	\and
             	Departamento de F\'{i}sica Te\'{o}rica, C-XI, Facultad de Ciencias, Universidad Aut\'{o}noma de Madrid, Cantoblanco, 28049 Madrid, Spain
	\and
		NASA Herschel Science Center, Infrared Processing and Analysis Center, MS 100-22, California Institute of Technology, Pasadena, CA 91125, USA
	\and
		Herschel Science Center - C11, European Space Agency (ESA), European Space Astronomy Centre (ESAC), P.O. Box 78, Villanueva de la Ca\~nada, 28691 Madrid, Spain
	\and
		European Southern Observatory, Casilla 1900, Santiago 19, Chile 
	\and
		Max Planck Institut f\"ur Astronomie, K\"onigstuhl 17, 69117 Heidelberg, Germany
	\and
	 	Astrophysics Science Division, NASA Goddard Space Flight Center, Greenbelt, MD 20771, USA
	\and
		Instituto Nacional de Astrof\'{\i}sica, \'Optica y Electr\'onica (INAOE), Aptdo. Postal 51 y 216, 72000 Puebla, Pue., Mexico		
          \and
         		UJF-Grenoble 1 / CNRS-INSU, Institut de Plan\'etologie et d'Astrophysique de Grenoble (IPAG) UMR 5274, Grenoble, F-38041, France 
          \and
          	Astrophysics Mission Division, Research and Scientific Support Department ESA, ESTEC, SRE-SA P.O. Box 299, Keplerlaan 1 NL-2200AG, Noordwijk, The Netherlands 
          \and
          	Astrophysikalisches Institut und Universit\"atssternwarte, Friedrich-Schiller-Universit\"at Jena, Schillerg\"a\ss chen 2-3, 07745 Jena, Germany 
           \and
		NASA Goddard Space Flight Center, Exoplanets and Stellar Astrophysics Laboratory, Code 667, Greenbelt, MD  20771, USA
	\and
		Observatoire de Paris, Section de Meudon 5, place Jules Janssen, 92195 MEUDON Cedex, Laboratoire d'\'etudes spatiales et d'instrumentation en astrophysique, France
	\and
		Dept. of Physics \& Astronomy, The Open University, Walton Hall, Milton Keynes MK7 6AA, UK
	\and
		Space Science \& Technology Department, CCLRC Rutherford Appleton Laboratory, Chilton, Didcot, Oxfordshire OX11 0QX, UK
            }

   \date{Received ; accepted }

 
  \abstract
   { Chromospheres and coronae are common phenomena on solar-type stars. Understanding the energy transfer to these heated atmospheric layers requires direct access to the relevant empirical data. Study of these structures has, by and large, been limited to the Sun thus far.}
   {The region of the temperature reversal can be directly observed only in the far infrared and submillimetre spectral regime. We aim at the determination of the characteristics of the atmosphere in the region of the  temperature minimum of the solar sister star \acena. This will  as a bonus also provide a detailed mapping of the spectral energy distribution, i.e. knowledge that is crucial when searching for faint, Kuiper belt-like dust emission around other stars.}
   {For the nearby binary system \acen, stellar parameters are known with high accuracy from measurements. For the basic model parameters $T_{\rm eff}$, $\log g$ and [Fe/H], we interpolate in the grid of GAIA/PHOENIX stellar model atmospheres and compute the corresponding model for the G2\,V star \acena. Comparison with photometric measurements shows excellent agreement between observed photospheric data in the optical and infrared. For longer wavelengths, the modelled spectral energy distribution is compared to {{\it Spitzer-}}MIPS, {{\it Herschel-}}PACS, {{\it Herschel-}}SPIRE and APEX-LABOCA photometry. A specifically tailored Uppsala model based on the MARCS code and extending further in wavelength is used to gauge the emission characteristics of \acena\  in the far infared.}
   {Similar to the Sun, the far infrared (FIR) emission of \acena\ originates in the minimum temperature region above the stellar photosphere in the visible. However, in comparison with the solar case, the FIR photosphere of \acena\ appears marginally cooler, $T_{\rm min}\sim T_{160\,\mu{\rm m}}= 3920 \pm 375$\,K. Beyond the minimum near 160\,\um, the brightness temperatures increase and this radiation likely originates in warmer regions of the chromosphere of \acena.}
   {To the best of our knowledge this is the first time a temperature minimum has been directly measured on a main-sequence star other than the Sun.}
   
    
    \maketitle
%

\section{Introduction}

\acen tauri is the nearest stellar system to the Sun, located at a distance of 1.3\,pc \citep[$\pi = 747.1 \pm 1.2$\,mas,][]{soderhjelm1999}. The physical binary is composed of two solar-like stars, the brighter of which, \acena\ (HIP 71683, HD 128620) a G2 V star, is often considered a ``solar twin" \citep{cayrel1996,melendez2009}. The companion, \acenb, is of slightly later spectral type (K1) and was recently found to host an Earth-mass planet \citep{dumusque2012}. A possible link between chemical composition in the atmospheres of solar twins and the formation of systems containing rocky planets was proposed by \citet{melendez2009}. 

No planet has yet been found around the primary, however, but like the Sun, \acena\ shows evidence for chromospheric emission in the optical and ultraviolet spectral regions \citep[][and references therein]{ayres1976,judge2004} and should therefore also have atmospheric regions where the temperature gradient turns from being negative to becoming positive. There, the radial temperature profile of the star should exhibit a minimum. The rise in temperature beyond the ``minimum temperature" region is caused by non-radiative energy being deposited which leads to the heating of the higher atmospheric levels. The responsible physical processes are as yet un-identified and constitute the subject of intense study of solar physics and stellar astrophysics \citep{kalkofen2007,wedemeyer2012,cohen2005,harper2012}. 

The minimum temperature of the solar atmosphere can be measured directly only in the far infared \citep{eddy1969,avrett2003}. In the wavelength region $\sim 50-350$\,\um, the atmosphere is becoming optically thick due to the dominating H$^{-}$ free-free opacity \citep[$\propto\!\lambda^2$;][]{geltman1965,doughty1966} and, consequently, radiates at lower temperatures than the layers beneath (the photosphere in the visible, where $\tau_{0.5\,\mu {\rm m}} >1$). The precise location of the temperature minimum is dependent on the detailed structure of the atmosphere and can, with no convincing theory at hand, be determined only from direct measurement. It is in this region, where non-radiative energy is deposited and its physics is of great general interest.

By analogy, this phenomenon could also be expected to be found on \acena. We therefore set out to obtain such data with the space observatory {\it Herschel} \citep{pilbratt2010} in the far infrared (FIR) and complementary ones with the ground-based APEX sub-millimetre (submm) telescope. These facilities allow photometric imaging observations with high sensitivity. We present our findings in this  {\it Letter}, which is organised as follows: in Sect.\,2, the observations and reduction of the data are briefly described. The results are presented and discussed in Sect.\,3, and in Sect.\,4, our main conclusions are summarised. 

\begin{table}
\caption{Photometry and radiation temperatures of \acen tauri A}             
\label{fluxes}      
\begin{tabular}{llcc}      
\hline\hline    
\noalign{\smallskip}             
$\lambda_{\rm eff}$	& $S_{\nu}$ 	&  $T_{\rm rad}$ &	Photometry	\\
(\um)				& (Jy)		& (K)			   &   \& Reference    \\
\noalign{\smallskip}	
\hline                        
\noalign{\smallskip}	
   0.440  	&  $2215 \pm 41$   	&  $5792  \pm   19$ 	&	B (1)    		\\
   0.550  	&  $3640 \pm 67$   	&  $5830  \pm   24$  	&	V (1)   		 \\
   0.790  	&  $4814 \pm 89$  	 &  $5775 \pm    32$	&	I  (1)   		 \\
   0.440  	&  $2356 \pm 43$   	&  $5856   \pm  19$	&	B  (2)               	 \\
   0.550  	&  $3606 \pm 66$  	&  $5818  \pm   23$	&	V  (2)               	\\
   0.640  	&  $4259 \pm 78$   	&  $5787   \pm  27$	&	R$_{\rm c}$ (2)	 \\
   0.790  	&  $4784 \pm 88$   	&  $5764 \pm    32$	&	I $\!_{\rm c}$ (2) \\
   1.215  	&  $4658 \pm 86$   	& $5928  \pm   47$	&	J   (3)       		 \\
   1.654  	&  $3744 \pm 69$   	& $6121  \pm   60$	&	H  (3)           	\\
   2.179  	&  $2561 \pm 47$   	& $5994   \pm  67$	&	K  (3)            	 \\
   3.547  	&  $1194 \pm 22$   	& $5856  \pm   78$	&	L  (3)            	 \\
   4.769  	&  $592   \pm 11$   	& $4995   \pm  69$	&	M  (3)       		\\
  24        	& $28.53 \pm 0.58$	& $4736   \pm  91$	&	MIPS (4)   	 	\\
  70       	&  $3.40 \pm 0.70$  	& $4599  \pm   937$	&	MIPS (4)     	 \\
  70       	&  $3.35 \pm 0.028$ & $4540   \pm  37$	&	PACS (5) 		\\
100      	&  $1.41 \pm 0.05$   &$3909  \pm  135$	&	PACS (6)		 \\
160      	&  $0.56 \pm 0.06$   &$3920  \pm 394$	&$^{\ast}$	PACS  (5), (6)	 \\
250	    	&  $0.24  \pm 0.05$  & $4084 \pm 845$    &$^{\ast}$SPIRE (5)	\\
350      	&  $0.145 \pm 0.028$& $4822 \pm 927$  &$^{\ast}$SPIRE (5)	\\
500		&   $0.08 \pm 0.03$   & $5421\pm 2018$ &$^{\ast}$SPIRE (5)	\\
870      	&  $0.028\pm 0.007$& $5738 \pm1432$ &$^{\ast}$LABOCA (7) \\
\noalign{\smallskip}	
\hline                        
\noalign{\smallskip}	
\end{tabular}
\begin{list}{}{}
\item[$^{\ast}$] Asterisks indicate values determined according to Eq.\,1.
\item[(1)] HIPPARCOS,  (2) \citet{bessell1990}, (3) \citet{engels1981}.
\item[(4)]  G.\,Bryden [priv. com.; FWHM(24\,\um) = 6\asec, (70\,\um) = 18\asec]. Binary separation on 9 April, 2005, \asecdot{10}{4}. 
\item[(5)] Hi-GAL: KPOT\_smolinar\_1, fields 314\_0 \& 316\_0. {\it Herschel}-beams FWHM(70\,\um) = \asecdot{5}{6}, (100\,\um) = \asecdot{6}{8}, (160\,\um) = \asecdot{11}{3}, (250\,\um) = \asecdot{17}{6}, (350\,\um) = \asecdot{23}{9}, (500\,\um) = \asecdot{35}{2}. Binary separation on 21 August, 2010, \asecdot{6}{3}. 
\item[(6)] DUNES: KPOT\_ceiroa\_1. Binary separation 29 July, 2011, \asecdot{5}{7}. 
\item[(7)] 384.C-1025, 380.C-3044(A): FWHM(870\,\um) = \asecdot{19}{5}. Binary separation 20-13 November, 2007, \asecdot{8}{8} and 19 September, 2009, \asecdot{7}{0}.
\end{list}
\end{table}

\begin{table}
\caption{Stellar parameters for \acena}             
\label{params}      
\begin{tabular}{lclcc}      
\hline\hline    
\noalign{\smallskip} 
Parameter, {\it P} (unit)		&\multicolumn{2}{c}{Value of $P$} 	& $(\Delta P/P)$	&  Ref.	\\
  						\cline{2-3}												
                                                                                    & Sun & \acena 	&\acena			&		\\
\hline          
\noalign{\smallskip}	
Mass, $M$ (\msun)						& 1.000 &1.105 	& 0.006			& (1)		\\
Radius, $R$ (\rsun)						&  1.000 &	1.224 	& 0.002			& (2)  	\\
Radiative luminosity, $L$ (\lsun)			& 1.000  &	1.549 	& 0.042			& (3)		\\
Effective temperature, $T_{\rm eff}$ 	(K)		& 5770 &	5824 	& 0.004			& (3) 	 \\
Surface gravity, $\log g$ 	(cgs)				& 4.440&	4.306 	& 0.001			& (3)		 \\
Metallicity, [Fe/H]						& 0.000& $+0.24$	& 0.166			& (3)		 \\
Age, $\mathcal{T}$ (Gyr)		& 4.57$^{\dagger}$	    &4.85 		& 0.103			&  (1)	 \\
\hline                                   
\end{tabular}
\begin{list}{}{}
\item[Ref.]  (1)  \citet{thevenin2002}, (2) \citet{kervella2003}, (3) \citet{torres2010}. $^{\dagger}$ \citet{bonanno2002} give $\mathcal{T}_{\odot}=4.57\pm 0.11$\,Gyr.
\end{list}
\end{table}

\section{Observations and data reduction}

On July 29, 2011, PACS scan maps \citep{poglitsch2010} of  \acen\ were obtained for the DUNES programme \citep{eiroa2013} at 100\,\um\ and 160\,\um\ at two array orientations (70\adeg\ and 110\adeg) to suppress detector striping. The selected scan speed was intermediate, i.e. 20\asec s$^{-1}$, determining the FWHM at the two wavelengths (\asecdot{6}{8} and \asecdot{11}{3}, respectively). In addition, PACS 70\,\um, 160\,\um\ and SPIRE 250\,\um, 350\,\um, 500\,\um\ \citep{griffin2010} data obtained as part of the Hi-GAL programme \citep{molinari2010}  were also analysed (see {\small \texttt{http://herschel.esac.esa.int/Docs/SPIRE/html/spire\_om.\\
html\#x1-980005.2.7}}). Our reduction work is thoroughly described by \citet{eiroa2013} and \citet{wiegert2013} and the PACS calibration scheme is reported in detail in {\small \texttt{http://herschel.esac.esa.int/twiki/bin/view/Public/Pacs\\ CalibrationWeb\#PACS\_instrument\_and\_calibration}}. Aperture corrections of the flux and sky noise corrections for correlated noise in the super-sampled images were done according to the technical notes PICC-ME-TN-037 and PICC-ME-TN-038 and {\small \texttt{https://nhscsci.ipac.caltech.edu/sc/index.php/Pacs/Ab-\\
soluteCalibration}}. Both the relative and absolute positions in the sky are well known \citep[][Table\,\ref{fluxes}]{pourbaix2002} and are shown, for the epochs of the observations, in Fig.\,1 of \citet{wiegert2013}.

The LABOCA \citep[][and \texttt{http://www.apex-\\
telescope.org/telescope/}]{siringo2009} observations of \acen\  were made during two runs, viz. in  2007, November 10--13, and in 2009, September 19. The data associated with the programmes 380.C-3044(A) and 384.C-1025(A) were retrieved from the ESO archive. The map data were reduced and calibrated using the software package CRUSH\,2 developed by Attila\,Kov\' acs, see {\small \texttt{http://www.submm.caltech.edu/$\sim$sharc/crush/download.htm}}. The data have been smoothed with a Gaussian of HPBW=13\asec, resulting in an effective FWHM=\asecdot{23}{4}. However, fluxes in Jy/beam are given for an FWHM=\asecdot{19}{5}.
 
All details regarding the data for both \acena\ and \acenb\ will be presented in a forthcoming paper \citep{wiegert2013}.

\section{Results and discussion}
 
 \subsection{The spectral energy distribution of \acena}

 \acen tauri was clearly detected at all observed wavelengths, with the possible exception at 500\,\um\ because of high background emission. Flux densities for \acena\ with their statistical error estimates, from 0.4 to 870\,\um, are provided in Table\,\ref{fluxes}, together with the references to the photometry. For data where the binary was spatially unresolved (indicated by asterisks in the table), we used the PHOENIX models for both A and B to estimate the relative flux contributed by \acena\  \citep{wiegert2013}, viz.  

\[ \frac {S_{\nu,\,{\rm A}} } {S_{\nu,\,{\rm A}} + S_{\nu,\,{\rm B}}}  \left  \{  	\begin{array} {ll}
                                         								           \ge 0.69\,\& \le 1.0 		& \mbox{if $\lambda \le 10\,\mu {\rm m}$}  \\
								 				           = {\rm const.} = 0.69         &  \mbox{if $\lambda > 10\,\mu {\rm m}$}\,\,.\hspace{2.2cm} (1)
											   			\end{array}
										  	       \right.  \] 
\setcounter{equation}{1}
This is in fair agreement with the PACS measurement at 70\,\um, where the pair is resolved and the flux ratio $S\!_{\rm A} /S\!_{\rm B}=2.45\pm0.45$, as compared to the PHOENIX value of 2.25. We are aware that using a constant value may not yield the correct answer at all wavelengths. However, with its deeper and more compact convection zone \acenb\ is the more active and more X-ray luminous star of the binary system \citep{dewarf2010}. Assuming that the wavelength of the temperature minimum is similar for both stars yields for \acena\ an optically determined $T_{\rm min}$ which is higher by 20\% \citep{ayres1976}. We will keep this caveat in mind when searching for the temperature minimum in \acena.
 
The observations of \acen\ are part of the DUNES programme \citep{eiroa2013} which focusses on nearby solar-type stars. The observations with {\it Herschel}-PACS \citep{poglitsch2010} at 100\,\um\ and 160\,\um\ are aiming at detecting the stellar photospheres at an $S/N\ge5$ at 100\,\um. Prior to {\it Herschel}, and surprisingly perhaps, not many data at long wavelengths and of high photometric quality are available for these stars. One reason likely being detector saturation issues  due to their brightness (e.g., WISE bands W1--W4), another due to contamination in the large beams by confusing emission near the galactic plane (e.g., IRAS, ISO-PHOT and AKARI data).  Another issue being how to define the photosphere at FIR wavelengths - which, in fact, is at the heart of this paper.

Being as close-by as 1.3\,pc and at comparable age, \acena\ is an excellent astrophysical laboratory for stars very similar to the Sun. From numerous literature sources, \citet{torres2010} compiled the currently best available basic stellar parameters and the estimated errors on the physical quantities are generally small (Table\,\ref{params}). However, whereas the tabulated uncertainty of the effective temperature of  \acena\ is less than half a percent, the observed spread in Table\,1 of \citet{porto2008} does correspond to more than ten times this much. On the other hand, the radius given by \citet{torres2010} is that directly measured by \citet{kervella2003} using interferometry (and corrected for limb darkening), with an error of 0.2\% \citep{bigot2006}. The mass has been obtained from asteroseismological measurements and is good to within 0.6\% \citep{thevenin2002}. However, it is also evident from the table that the twinship of \acena\ and the Sun is not identically close in every detail.

For such an impressive record of accuracy for the stellar parameters of the \acen\ components it should be possible to construct theoretical model photospheres with which observations can be directly compared to a high level of precision. In this paper, we report on FIR and submm observations, which should provide valuable constraints on the Spectral Energy Distribution (SED). This could potentially be useful to gauge the temperature minimum at the base of the stellar chromosphere. A clear understanding of the latter is crucial when attempting to determine extremely low levels of cool circumstellar dust emission \citep{eiroa2011, eiroa2013}. Because of binary dynamics, dust is not expected to contribute to the stellar radiation from \acena\ \citep[][and references therein]{wiegert2013}. 

\begin{figure*}
  \resizebox{\hsize}{!}{
    \rotatebox{00}{\includegraphics{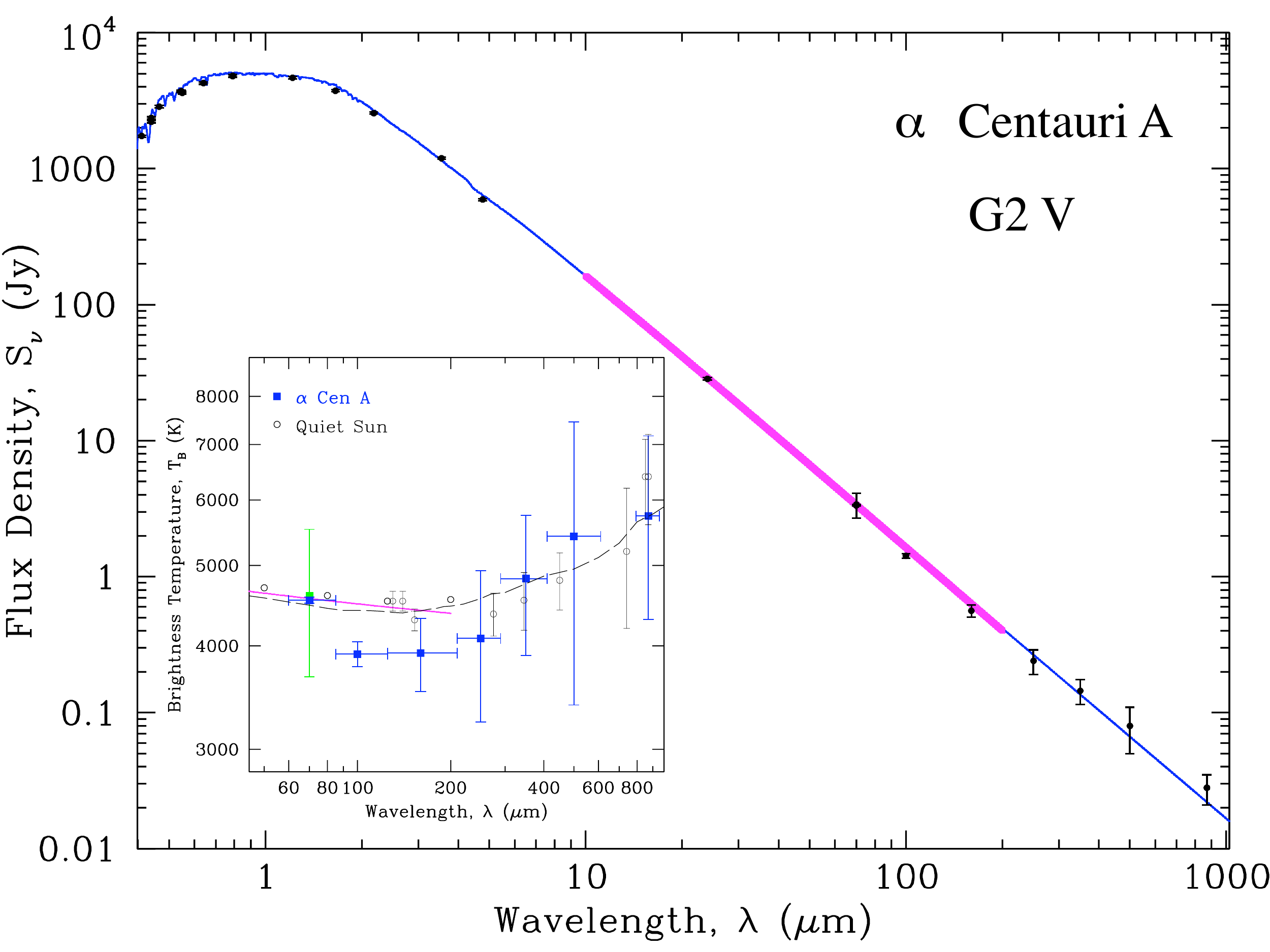}}
                        }
  \caption{The SED of \acena, where the blue line represents the PHOENIX model photosphere (computed up to $\lambda = 45$\,\um\ and Rayleigh-Jeans extrapolated beyond) and the thick (purple) line the Uppsala model photosphere (extending to  $\lambda = 200$\,\um). Photometric data are shown with their $1 \sigma$ error bar estimates (Table\,\ref{fluxes}). The inset shows the brightness temperature of \acena\ (squares)  and the Quiet Sun (circles) in the far infrared and submm. The solar data from the compilations by \citet{gu1997} and \citet{loukitcheva2004} are shown together with a semi-empirical chromosphere model for the Sun \citep[][black dashes: VAL\,IIIC]{vernazza1981}.  At 70\,\um, the symbol with the larger error bar (green) represents the \ {\it Spitzer}-MIPS datum and filled (blue) squares {\it Herschel}-PACS/SPIRE and LABOCA data. Horisontal bars indicate filter widths. 
   }
  \label{SED}
\end{figure*}

The atmosphere model for \acena\ has been computed by a 3D interpolation in the high-resolution PHOENIX grid for GAIA \citep{brott2005} and with the following parameters: ($T_{\rm eff},\,\log g,\, {\rm [Fe/H]}$)=(5824\,K, 4.306, +0.24), where [Fe/H]$_{\star}=\log (N_{\rm Fe}/N_{\rm H})_{\star} - \log (N_{\rm Fe}/N_{\rm H})_{\odot}$. This model is shown in Fig.\,\ref{SED}, together with the excellently agreeing photometry (Table\,\ref{fluxes}). The photometry has not been used in the analysis, but serves to illustrate the good agreement between this model and the observations.  

The PHOENIX model spectra are computed up to 45\,\um. These models are used extensively by the DUNES programme and therefore also exploited here. However, for this particular study, we have also used a specifically tailored model (the ``Uppsala model") based on the MARCS code \citep{gustafsson2008}, with, of course, exactly the same atmosphere parameters as before. This model computation extends to atmospheric layers more than 1500\,km up, including those regions which emit  predominantly around 200\,\um. The infrared part of the Uppsala model is shown superposed onto the PHOENIX model in Fig.\,\ref{SED}, where it can be seen that these two atmosphere models are virtually indistinguishable.

\subsection{The temperature minimum of \acena}

The observed FIR fluxes of \acena\ seem somewhat lower than in the model but appear to turn upward in the submm/mm bands (inset of Fig.\,\ref{SED}). The 1D Uppsala (or PHOENIX) model describes stellar atmospheres in local thermodynamic equilibrium (LTE) and does not account for any temperature inversions. Both non-LTE effects and the change of the temperature gradient will influence the emergent spectrum \citep[for details, see][]{delaluz2011}. In the far infrared, this emission from the solar chromosphere exhibits a minimum in brightness temperature around 150\,\um\  \citep[][and references therein]{avrett2003}.  A chromosphere has also been observed for \acena\ in ultraviolet line emission \citep[e.g.,][]{judge2004}. From the interpretation of the wings of the optical CaII\,K line, \citet{ayres1976} deduced a value of $T_{\rm min}/T_{\rm eff}=0.78$, similar to the solar value of 0.77. According to \citet{judge2004}, the level of activity of \acena\ is low and similar to that of the Sun in an intermediate state of its cycle and, on the basis of long-time X-ray monitoring, this apparent lack of variability was also emphasised by \citet{ayres2009}.

We assume an atmosphere in radiative and hydrostatic equilibrium. The flux received at the Earth, $S\!_{\nu}$, is related to the outward flux through the surface of the star, $F\!_{\nu}$ \citep[e.g.,][]{mihalas1978}, through

\begin{equation}
S_{\nu} = \pi \left ( \frac {\phi_{\nu}}{2} \right )^2 \frac {F\!_{\nu}}{\pi}
\end{equation}

where $\phi_{\nu}$ is the angular diameter of the stellar disc of radius $R_{\nu}$, at a given frequency $\nu$ or wavelength $\lambda$, and as seen at a distance $D$. For these parallel stellar rays, a brightness temperature, $T_{\rm B}(\nu)$, can be defined through the Planck function by

\begin{equation}
F\!_{\nu} = \pi\,B_{\nu}(T_{\rm B})\,\,,
\end{equation}

so that with $\phi_{\nu}/2 = R_{\nu}/D$,

\begin{equation}
T_{\rm B}(\nu) = \frac{h\,\nu}{k} \left [  \ln \left ( \frac{2\,\pi\,R_{\nu}^2\,h\,\nu^3}{D^2\,c^2\,S\!_{\nu}}  + 1\right ) \right ]^{-1}\,\,.
\end{equation}

In the far infrared, $R_{\rm FIR}=R_{0.5} + \Delta R$, where $R_{0.5}$ is the ``photospheric" radius of the star. There the optical depth in the visible is unity ($\tau_{0.5}\sim1$). The temperature minimum is found in regions higher up, where $\tau_{\rm FIR} > 1$ (but  $\tau_{0.5}\ll 1$). It is straightforward to show \citep[e.g.,][]{tatebe2006} that $\Delta R/R_{\rm FIR}$ is of the order of a few times \powten{-4}. Therefore, $\Delta R$ corresponds to some 500\,km, and will per se not introduce any significant errors for stars like the Sun and \acena\ (luminosity class V) and in Eq.\,3, we use $R_{\rm FIR}=R_{0.5}$ (cf. Table\,\ref{params}). However,  over this distance, the density drops by an order of magnitude, making model computations increasingly difficult.

In Fig.\,\ref{SED}, $T_{\rm B}(\nu)$ is shown for the Uppsala model photosphere, where LTE is assumed for the free-free H$^{-}$ continuum, together with a semi-empirical chromospheric model of the quiet Sun  
\citep[VAL\,IIIC,][]{vernazza1981}. Solar data are shown as open circles and the observations of \acena\ as filled squares. 

Using the Uppsala model, we find that $\chi_{100\,\mu \rm m}= - 4.4$, where $\chi_{\nu} = (S_{\nu,\,\rm obs} - S_{\nu,\,\rm mod})/\sigma_{\nu}$. The difference in brightness temperature by 500\,K in the 100 to 160\,\um\ region could therefore be significant and it cannot be excluded that in \acena\ the atmosphere becomes optically thick at higher, cooler levels than the model and that the structure is different from that in the Sun\footnote{Actually, a very similar scenario is proposed also for the Sun by \citet{ayres2002} in order to explain the lower $T_{\rm min}$ indicated by observations of infrared CO lines.}. Important is however the fact that the FIR-SED indeed goes through a minimum, as the brightness temperatures rise at longer wavelengths and, at e.g. 870\,\um, H free-free emission from the stellar chromosphere appears to dominate. 

In the solar atmosphere, the temperature minimum occurs around 150\,\um. This seems to be the case also for \acena, where $T_{160\mu \rm m} = 3920 \pm 375$\,K. It is customary to express the minimum temperature of the stellar atmosphere in terms of the effective temperature, $T_{\rm eff}$, and for \acena, $T_{\rm min}/T_{\rm eff} = 0.67\pm0.06$. \citet{ayres1976} had earlier, from Ca\,II K-line fitting, estimated this ratio as 0.78 to 0.79, for an assumed $T_{\rm eff}= 5770$ to 5700\,K, hence  $T_{\rm min} \sim 4500$\,K, which would indicate an ``optical minimum temperature" about 500\,K higher than what we determined from a direct measurement in the far infrared. This result may seem surprising, but we have to recall that similar is seen in the Sun, where different diagnostics yield a span in minimum temperature of about 600\,K \citep{avrett2003}.  It is also worth remembering that even for the Sun, a unique model for its chromospheric emission has yet to be found \citep[e.g.,][]{delaluz2011}. For its ``sister star" \acena, future data at submm and mm wavelengths, with for instance ALMA, should contribute to a better characterisation of the atmosphere of our nearest neighbour.

\section{Conclusions}

\begin{itemize}
\item[$\bullet$] We successfully observed the far infrared energy distribution of the nearby G2\,V star \acena\ with instruments aboard {\it Herschel} and from the ground.
\item[$\bullet$] The observed radiation temperatures appear lower than what is expected on the basis of extensions of LTE stellar atmosphere models. Near 160\,\um, the minimum temperature in the atmosphere of \acena, $T_{\rm min}/T_{\rm eff}=0.67\pm 0.06$, is marginally lower than the ratio of 0.73 observed in the Sun at 150\,\um. 
\item[$\bullet$] At 870\,\um, however, the emission from \acena\ appears to originate from regions at higher temperatures, as might be expected, viz. $T_{870\,\mu \rm m}/T_{\rm eff}=1.0\pm 0.2$.
\item[$\bullet$] Cold temperature minimum regions will lead to the underestimation of the amount of dust present in cold debris discs.
\end{itemize}

\begin{acknowledgements} 
We thank Dr. K.\,Eriksson for the special computations of the \acena-Uppsala model. The Swedish authors appreciate the continued support by the Swedish National Space Board (SNSB) for our {\it Herschel}-projects. C.\,Eiroa, J.P.\,Marshall and B.\,Montesinos are partially supported by Spanish grant AYA 2011/02622. A.\,Bayo was co-funded under the Marie Curie Actions of the European Comission (FP7-COFUND). S.\,Ertel thanks the French National Research Agency (ANR) for financial support through contract ANR-2010 BLAN-0505-01 (EXOZODI). 
\end{acknowledgements}

\begin{thebibliography}{}
\bibitem[Avrett(2003)]{avrett2003} Avrett E.H., 2003, ASPCS, 286, 419
\bibitem[Ayres (2009)]{ayres2009} Ayres T.R., 2009, ApJ, 696, 1931
\bibitem[Ayres et al.(1976)]{ayres1976} Ayres T.R., Linsky J.L., Rodgers A.W. \& Kurucz R.L., 1976, ApJ, 210, 199
\bibitem[Ayres(2002)]{ayres2002} Ayres T.R., 2002, ApJ, 575, 1104
\bibitem[Bessell(1990)]{bessell1990} Bessell M.S., 1990, PASP, 102, 1181
\bibitem[Bigot et al.(2006)]{bigot2006} Bigot L., Kervella P., Th\' evenin F. \& S\' egransan D., 2006, A\&A, 446, 635
\bibitem[Bonanno et al.(2002)]{bonanno2002} Bonanno A., Schlattl H. \& Patern\`o L., 2002, A\&A, 390, 1115
\bibitem[Brott \& Hauschildt(2005)]{brott2005} Brott I. \& Hauschildt P.H., 2005, ESASP, 576, 565
\bibitem[Cayrel de Strobel(1996)]{cayrel1996} Cayrel de Strobel G., 1996, A\&A Rev., 7, 243
\bibitem[Cohen et al.(2005)]{cohen2005} Cohen M., Carbon D.F., Welch W.J., et al., 2005, AJ, 129, 2836
\bibitem[De la Luz et al.(2011)]{delaluz2011} De la Luz V., Lara A. \& Raulin J.-P., 2011, ApJ, 737, 1
\bibitem[DeWarf et al.(2010)]{dewarf2010} DeWarf L.E., Datin D.M. \& Guinan E.F., 2010, ApJ, 772, 343
\bibitem[Doughty \& Fraser(1966)]{doughty1966} Doughty N.A. \& Fraser P.A., 1966, MNRAS, 132, 267
\bibitem[Dumusque et al.(2012)] {dumusque2012} Dumusque X., Pepe F., Lovis C., et al. 2012, Nat, 491,, 207
\bibitem[Eddy et al.(1969)]{eddy1969} Eddy J.A., L\' ena P.J. \& MacQueen R.M., 1969, Sol.Phys., 10, 330
\bibitem[Eiroa et al.(2011)]{eiroa2011} Eiroa C., Marshall J.P., Mora A., et al., 2011, A\&A, 536, L4
\bibitem[Eiroa et al.(2013)]{eiroa2013} Eiroa C., et al.,  2013, in preparation
\bibitem[Engels et al.(1981)]{engels1981} Engels D., Sherwood W.A., Wamsteker W. \& Schultz G.V., 1981, A\&AS, 45, 5
\bibitem[Geltman(1965)]{geltman1965} Geltman S., 1965, ApJ, 141, 376
\bibitem[Griffin et al.(2010)]{griffin2010} Griffin M.J., Abergel A., Abreu A., et al., 2010, A\&A, 518, L3
\bibitem[Gu et al.(1997)] {gu1997} Gu Y., Jefferies J.T., Lindsay C. \& Avrett E.H., 1997, ApJ, 484, 960
\bibitem[Gustafsson et al.(2008)] {gustafsson2008} Gustafsson B., Edvardsson B., Eriksson K., et al., 2008, A\&A, 486, 951
\bibitem[Harper et al.(2012)]{harper2012} Harper G.M., O'Riain N. \& Ayres T.R., 2012, astro-ph 1210.2627
\bibitem[Judge et al.(2004) ]{judge2004}Judge P.G., Saar S.H., Carlsson M. \&  Ayres T.,R., 2004, ApJ, 609, 392
\bibitem[Kalkofen(2007)]{kalkofen2007} Kalkofen W., 2007, ApJ, 671, 2154 
\bibitem[Kervella et al.(2003)]{kervella2003} Kervella P., Th\' evenin F., S\' egransan D. et al., 2003, A\&A, 404, 1087
\bibitem[Loukitcheva et al.(2004)]{loukitcheva2004} Loukitcheva M., Solanki S.K., Carlsson M. \& Stein R.F., 2004, A\&A, 419, 747
\bibitem[Mel\'endez et al.(2009)]{melendez2009} Mel\'endez J., Asplund M., Gustafsson B. \& Yong D., 2009, ApJ, 704, L66
\bibitem[Mihalas(1978)]{mihalas1978} Mihalas D., 1978, Stellar Atmospheres, W.H. Freeman
\bibitem[Molinari et al.(2010)]{molinari2010} Molinari S., Swinyard B., Bally J., et al., 2010, A\&A, 518,  L100
\bibitem[Pilbratt et al.(2010)]{pilbratt2010} Pilbratt G.L., Riedinger J.R., Passvogel T., et al., 2010, A\&A, 518, L1 
\bibitem[Poglitsch et al.(2010)]{poglitsch2010} Poglitsch A., Waelkens C., Geis N., et al., 2010, A\&A, 518, L2
\bibitem[Porto de Mello et al.(2008)] {porto2008} Porto de Mello G.F., Lyra W. \& Keller G.R., 2008, A\&A,488, 653
\bibitem[Pourbaix et al.(2002)]{pourbaix2002} Pourbaix D., Nidever D., McCarthy C., et al., 2002, A\&A, 386, 280
\bibitem[Siringo et al.(2009)]{siringo2009} Siringo G., Kreysa E., Kov\' acs A., et al., 2009, A\&A, 497, 945
\bibitem[S\"oderhjelm(1999)]{soderhjelm1999} S\"oderhjelm S., 1999, A\&A, 341, 121
\bibitem[Tatebe \& Townes(2006)]{tatebe2006} Tatebe K. \& Townes C.H., 2006, ApJ, 644, 1145
\bibitem[Th\' evenin et al.(2002)]{thevenin2002} Th\' evenin F., Provost J., Morel P., et al., 2002, A\&A, 392, L9
\bibitem[Torres et al.(2010)]{torres2010} Torres G., Andersen J. \& Gim\' enez A., 2010, A\&A Rev., 18, 67 
\bibitem[Vernazza et al.(1981)]{vernazza1981} Vernazza J.E., Avtrett E.H. \& Loeser R., 1981, ApJS, 45, 635
\bibitem[Wedemeyer$-$B\"ohm S., et al.(2012)]{wedemeyer2012} Wedemeyer-B\"ohm , Nature 486, 505
\bibitem[Wiegert et al.(2013)]{wiegert2013} Wiegert J., et al., 2013, in preparation
\end{thebibliography}
\end{document}